\documentclass[onecolumn,preprint,preprintnumbers,amsmath,amssymb]{revtex4}

\usepackage{graphicx}% Include figure files
\usepackage{dcolumn}% Align table columns on decimal point
\usepackage{bm}% bold math

\begin{document}

\title{Electromagnetic focusing through a tilted dielectric surface}

\author{Lars Egil Helseth}
\address{Max Planck Institute of Colloids and Interfaces, D-14424 Potsdam, Germany}%
\altaffiliation[Also at ]{Department of Physics, University of Oslo,  P.O. Box 1048 Blindern, N-0316 Oslo, Norway. \\
Email: l.e.helseth@fys.uio.no}%

\begin{abstract}
Focusing through a tilted dielectric interface is studied, and an explicit expression for 
the electric field in the focal region is found. In the case of small tilt angles, only 
a simple aberration term remains.
\end{abstract}

\maketitle

\section{Introduction}      
Focusing of light has drawn considerable attention for many centuries due to
its importance in many fields of physics. Microscopy, spectroscopy and 
optical data storage are only a few of the fields utilizing focused light. 
The earliest treatment of electromagnetic focusing problems are due to
Ignatowsky\cite{Ignatowsky}.  
However, the structure of the focused field was not clarified until the 1950's 
and 60's, when Wolf and coworkers did the first detailed studies of aplanatic 
electromagnetic focusing systems\cite{Wolf,Richards,Boivin}. Here the socalled Debye 
approximation was adopted, where only the plane waves with propagation vectors 
that fall inside the geometrical cone whose apex is at the focal point 
contribute to the field in the focal region\cite{Wolf,Debye}. The Debye
approximation predicts that the electromagnetic field is symmetrical about the
focal plane. However, later it was found that the Debye theory is valid only when 
the focal point is located many wavelengths away from the aperture, the angular 
aperture is sufficiently large, and the wavefront deviation kept reasonably 
low\cite{Wolf1,Stamnes,Stamnes1,Sheppard,Li}. In the general case the electromagnetic field
is not symmetrical about the geometrical focal plane, and focal shifts may occur.

Focusing through dielectric interfaces is of broad interest since this
geometry is often used in optical data storage and biology (see e.g. 
Refs. \cite{Stamnes2,Dhayalan,Sheppard1,Wilson,Torok,Ando,Hendriks,Helseth} and references therein). 
To date, most studies have concentrated about focusing through a dielelectric interface with its
normal along the optical axis. However, in some applications it is of importance to understand 
what happens when the interface is tilted. For example, consider an optical disk which rotates 
around its axis in a DVD player. If the disk is slightly tilted (e.g. by shaking the DVD player), 
then the wavefront is aberrated, and the focus is distorted. To find out how much the focus is 
distorted, it is most common to perform optical ray tracing or use scalar wave theory 
(see e.g. Refs. \cite{Ando,Hendriks} and references therein). However, also in the case of 
electromagnetic focusing it is of interest to obtain an explicit expression for the intensity 
distribution in the focal region, since this may give a clearer understanding of the physical 
problem. The aim of this Letter is to treat the problem of electromagnetic focusing through a tilted 
dielectric interface in more detail. First I will formulate the problem, and then 
look at the small angle limit in the Debye approximation. It should be emphasized that no 
numerical analysis is given here.

\section{General formulation}

In general, the diffracted field near the focal plane can be calculated 
in the Kirchoff approximation as \cite{Stamnes2} 
\begin{equation}
\mbox{\boldmath $E$} = -\frac{ik_{i}}{2\pi} \int_{-\infty}^{\infty} \int_{-\infty}^{\infty}
\mbox{\boldmath $T$} (\mbox{\boldmath $s$}_{i})
\exp[ik_{i}(s_{ix}x+s_{iy}y+s_{iz}z)] ds_{ix}ds_{iy} \,\,\, ,
\label{a}
\end{equation}
where $k_{i}=2\pi n_{i}/\lambda $ 
is the wavenumber in medium i, $\mbox{\boldmath$s$}_{i}=(s_{ix},s_{iy},s_{iz})$ is the 
unit vector along a typical ray, and $\mbox{\boldmath$T$}$ the vector 
pupil distribution which accounts for the polarization, phase and amplitude 
distributions at the exit pupil. 
Let us consider focusing through a tilted isotropic dielectric interface when the 
optical axis crosses the interface at a distance $z_{i}=-d$ (see Fig. 1). 
Without loss of generality, we let the angle $\beta$ be the tilting angle about the y-plane. 
Thus, the z-coordinate at the interface is given by $z_{i}=-d+xtan\beta $. When $\beta =0$ we 
have $\mbox{\boldmath$s$}_{1}=(s_{1x},s_{1y},s_{1z})$. The unit vector corresponding to a 
finite tilt angle is then given by a rotation of the coordinate system 
$\mbox{\boldmath$s$}_{1}'=\mbox{\boldmath$Ks$}_{1}$, where
\begin{displaymath}
\mathbf{K} = 
\left[ \begin{array}{ccc} 
cos\beta & 0 & -sin\beta \\
0 & 1 & 0 \\
sin\beta       & 0   & cos\beta   \\
\end{array} \right] \,\,\,  ,
\end{displaymath}
and $\mbox{\boldmath$s$}_{1}'=(s_{1x}cos\beta -s_{1z}sin\beta ,s_{1y},s_{1x}sin\beta +s_{1z}cos\beta)$.    
Here we have used the equivalence between tilting the optical axis and the dielectric interface for 
a single ray. 
We will see that this change of coordinate system is useful for expressing the electric field in the 
second medium. Moreover, Snell's law is now expressed as; $n_{1}s_{1y}=n_{2}s_{2y}$ 
and $n_{1}(s_{1x}cos\beta -s_{1z}sin\beta) =n_{2}s_{2x}$. We note that 
these equations are not valid when $n_{1}=n_{2}$ and $\beta \neq 0$, since in this case 
we can not immediately assume equivalence between tilting the (absent) substrate and the 
optical axis.

It is convenient to write the electric fields in the first and second media as (see also 
Ref. \cite{Torok})
\begin{equation}
\mbox{\boldmath $E$}_{1} =-\frac{ik_{1}}{2\pi} \int_{-\infty}^{\infty} \int_{-\infty}^{\infty}
\frac{\mbox{\boldmath $T$} (\mbox{\boldmath $s$}_{1})}{s_{1z}}
\exp[ik_{1}(s_{1x}x+s_{1y}y+s_{1z}z)] ds_{1x}ds_{1y} \,\,\, ,
\end{equation}
and
\begin{equation}
\mbox{\boldmath $E$}_{2} =-\frac{ik_{2}}{2\pi} \int_{-\infty}^{\infty} \int_{-\infty}^{\infty}
\mbox{\boldmath $T$} (\mbox{\boldmath $s$}_{2})
\exp[ik_{2}(s_{2x}x+s_{2y}y+s_{2z}z)] ds_{2x}ds_{2y} \,\,\, .
\end{equation}
To express the electric field in medium 2 in terms of 'untilted' 
coordinates, $(s_{1x},s_{1y})$, we note that
\begin{equation}
ds_{2x}ds_{2y}= J ds_{1x}ds_{1y} \,\,\, ,
\end{equation}
where the Jacobian determinant is given by 
\begin{equation}
J =\left( \frac{k_{1}}{k_{2}} \right) ^{2} \left( cos\beta + 
\frac{s_{1x}}{s_{1z}} sin\beta \right) \,\,\, .
\end{equation}
Moreover, we must match the electric field at the interface, which gives
\begin{equation}
\mbox{\boldmath $T$}_{2} =\mbox{\boldmath $F$}(\mbox{\boldmath $s$}_{1},\beta)  \frac{k_{1}\mbox{\boldmath $T$} (\mbox{\boldmath $s$}_{1})}{Jk_{2}s_{1z}} \exp\left[ iz_{i}(k_{1}s_{1z}-k_{2}s_{2z}) \right]\,\,\, ,
\end{equation}
where $\mbox{\boldmath $F$}(\mbox{\boldmath $s$}_{1},\beta)$ is the transmission factor (Fresnel coefficients) 
through the tilted surface. Now we may write
\begin{equation}
\mbox{\boldmath $T$}_{2} =\mbox{\boldmath $\tilde{T}$} (\mbox{\boldmath $s$}_{1},\beta) \exp(i\Psi _{M} +i\Psi _{T} ) \,\,\, ,
\end{equation}
where
\begin{equation}
\mbox{\boldmath $\tilde{T}$} (\mbox{\boldmath $s$}_{1},\beta)= 
\frac{k_{2}}{k_{1}} \frac{\mbox{\boldmath $F$}(\mbox{\boldmath $s$}_{1},\beta)  \mbox{\boldmath $T$} (\mbox{\boldmath $s$}_{1}) }{\left( cos\beta + 
\frac{s_{1x}}{s_{1z}} sin\beta \right)s_{1z} } \,\,\, .
\end{equation} 
$\Psi _{M}$ and $\Psi _{T}$ are given by 
\begin{equation}
\Psi _{M} =-d(k_{1}s_{1z} -k_{2}s_{2z}) \,\,\, ,
\end{equation} 
and
\begin{equation}
\Psi _{T} =xtan\beta (k_{1}s_{1z} -k_{2}s_{2z}) \,\,\, .
\end{equation}
Naturally, $\Psi _{T}$ vanishes when there is no index mismatch.  
Note also that $s_{2z}=\sqrt{1-s_{2x}^{2}-s_{2y}^{2}}$, where $s_{2x}$ and $s_{2y}$ 
are given by Snells law. 
Finally, the expression for the electric field in medium 2 is
\begin{equation}
\mbox{\boldmath $E$}_{2} = -\frac{ik_{2}^{2}}{2\pi k_{1}} \int_{-\infty}^{\infty} \int_{-\infty}^{\infty}
\mbox{\boldmath $T$}_{2} \exp \left[ ik_{1}(s_{1x}cos\beta -s_{1z}sin\beta )x + ik_{1}s_{1y}y + ik_{2}s_{2z}z \right] ds_{1x}ds_{1y} \,\,\, .
\label{Kir}
\end{equation}
This is expression enable us to calculate the electric field for reasonable tilt angles in
the Kirchoff approximation. In the next section I will study the special case of high angular 
apertures and small tilt angles, and it will be shown that only a simple aberration term remains.  

\section{Small tilt angles in the Debye approximation}
We now consider focusing with high angular aperture assuming that the Debye approximation can be used.
Then only the plane waves with propagation vectors 
that fall inside the geometrical cone whose apex is at the focal point 
contribute to the field in the focal region. Moreover, we assume that 
$\beta \ll 1$ and $z=0$. Then the influence of $\beta$ on 
$\tilde{\mbox{\boldmath ${T}$}} (\mbox{\boldmath $s$}_{1},\beta)$ can be neglected, and only 
the lowest order contribution in the phase remains. Equation (\ref{Kir}) can therefore be 
written as 
\begin{equation}
\mbox{\boldmath $E$}_{2} =-\frac{ik_{2}^{2}}{2\pi k_{1}} \int \int_{\Omega}
\tilde{\mbox{\boldmath $T$}} (\mbox{\boldmath $s$}_{1}) \exp(i\Psi _{T}' +i\Psi _{M}') 
\exp \left[ ik_{1}s_{1x} (x-u) + ik_{1}s_{1y}y \right] ds_{1x}ds_{1y} \,\,\, ,
\label{smalltilt}
\end{equation}
where $\Omega$ is the solid angle formed by all the geometrical rays,
\begin{equation}
\Psi _{T}' = -x\beta k_{2}s_{2z0} \,\,\, ,
\end{equation}
\begin{equation}
\Psi _{M}' = -dk_{1}s_{1z} + dk_{2}s_{2z0} \,\,\, ,
\end{equation}
$u=d\beta (k_{1}/k_{2})(s_{1z}/s_{2z0})$ and 
$s_{2z0}= \sqrt{1-(n_{1}/n_{2})^{2}(s_{1x}^{2} +s_{1y}^{2})}$. These expression were found by 
expanding $s_{2z}$ and only keeping the lowest order in $\beta$. Note that $\Psi _{M}'$ is the 
usual aberration introduced by focusing through a planar, nontilted dielectric surface, see 
Ref. \cite{Torok}. This aberration term will be neglected here, which is a reasonable approximation 
for systems corrected for the index mismatch (in absence of tilting). 
It is seen that the tilt introduces a coordinate shift, u, which depends on the tilt and index mismatch.
Note in particular that $u\approx d\beta $ when $n_{1} \approx n_{2}$, which is just the shift 
expected from a tilt in the coordinate system. Moreover, u is clearly altered 
when the index mismatch and numerical aperture change. In the further 
studies I will neglect this shift, concentrating on the remaining aberration term $\Psi _{T}'$. 
This term is also nonzero when $n_{1} \approx n_{2}$. However, I argue that it has the correct symmetry, 
and is a real aberration term due to the finite tilt $\beta$.

I now assume that the aperture is circular symmetric, which means that it is most convenient 
to adopt spherical coordinates:
\begin{equation}
\mbox{\boldmath$s$}_{i}=(sin\theta _{i} cos\phi,sin\theta _{i}sin\phi ,cos\theta _{i}) \,\,\, ,
\end{equation}
and
\begin{equation}
\mbox{\boldmath$r$}_{c}=(r_{c}sin\theta _{c} cos\phi _{c} ,r_{c} sin\theta _{c} sin\phi _{c}  ,z) \,\,\, .
\end{equation}
Equation (\ref{smalltilt}) can now be written as
\begin{eqnarray}
\mbox{\boldmath $E$}_{2} & = & -\frac{ik_{2}^{2}}{2\pi k_{1}} \int_{0}^{\alpha} \int_{0}^{2\pi}
\tilde{\mbox{\boldmath $T$}} (\theta _{1},\phi) \exp(-ik_{2}s_{2z0}\beta r_{c}sin\theta _{c}cos\phi _{c} ) \\
& &\exp \left[ ik_{1}r_{c}sin\theta _{1}sin\theta _{c} sin\theta _{1}cos(\phi -\phi _{c} \right)] d\phi d\theta _{1} \,\,\, ,
\end{eqnarray}

In order to be able to evaluate this integral, it remains to find the vector pupil function 
$\tilde{\mbox{\boldmath $T$}}(\theta _{1},\phi)$. This can be done following the guidelines given in e.g. 
Ref. \cite{Helseth}, which results in 
\begin{displaymath}
\tilde{\mathbf{T}}(\theta _{1},\phi) = A(\theta _{1})
\left[ \begin{array}{ccc} 
a [t_{p} cos\theta _{2} cos^{2}\phi +t_{s}sin^{2}\phi ]
+b[t_{p}cos\theta _{2}sin\phi cos\phi - t_{s}sin\phi cos\phi ] \\
a [t_{p}cos\theta _{2}cos\phi sin\phi - t_{s}sin\phi cos\phi ] +
b [t_{p}cos\theta _{2}sin^{2}\phi +t_{s}cos^{2}\phi ] \\
- t_{p}sin\theta _{2}[acos\phi + bsin\phi ]\\ 
\end{array} \right] \,\,\, ,
\label{q}
\end{displaymath}
where $A(\theta _{1})$ is an apodization factor, $a(\theta _{1},\phi) $ is the strength of the incident 
x polarized light, $b(\theta _{1},\phi)$ the strength of the incident y polarized light and $t_{p,s}$ the 
Fresnel transmission coefficients.
In the case of linearly polarized light (a=1, b=0) we obtain the following electric field components:
\begin{equation}
E_{x}\propto i\left( I_{0} +I_{2}cos2\phi_{c} \right) \,\,\, ,
\end{equation}
\begin{equation}
E_{y}\propto iI_{2}sin2\phi_{c} \,\,\, ,
\end{equation}
\begin{equation}
E_{z}\propto 2I_{1}cos\phi_{c} \,\,\, ,
\end{equation} 
where 
\begin{equation}
I_{0}=\int_{0}^{\alpha }
A(\theta _{1})(t_{s} +t_{p}cos\theta _{2})sin\theta _{1}J_{0}(kr_{c}sin\theta _{1}sin\theta_{c} )
\exp(-ik_{2}s_{2z0}\beta r_{c}sin\theta _{c}cos\phi _{c} )d\theta _{1}\,\,\, ,
\end{equation}
\begin{equation}
I_{1}=\int_{0}^{\alpha } 
A(\theta _{1})t_{p}sin\theta _{1} sin\theta _{2}J_{1}(k r_{c}sin\theta _{1}sin\theta_{c} )
\exp(-ik_{2}s_{2z0}\beta r_{c}sin\theta _{c}cos\phi _{c} )d\theta _{1}\,\,\, ,
\end{equation}
\begin{equation}
I_{2}= \int_{0}^{\alpha } 
A(\theta _{1})(t_{s} -t_{p}cos\theta _{2})sin\theta _{1}J_{2}(kr_{c}sin\theta _{1}sin\theta_{c} )
\exp(-ik_{2}s_{2z0}\beta r_{c}sin\theta _{c}cos\phi _{c} )d\theta _{1}\,\,\, .
\end{equation}
We see that now the only effect of the tilt is to 
introduce an aberration term which depends on the azimuthal angle $\phi _{c}$ as well 
as $\theta _{1}$. When $\beta =0$, these equations reduces to the ones found in 
e.g. Refs. \cite{Torok,Helseth}. 

In conclusion, focusing through a tilted dielectric interface has been investigated. It is 
found that the tilt introduces additional aberration terms. In the small angle limit and Debye 
approximation, only a simple aberration term remains.

\newpage
\begin{figure}
\includegraphics[width=12cm]{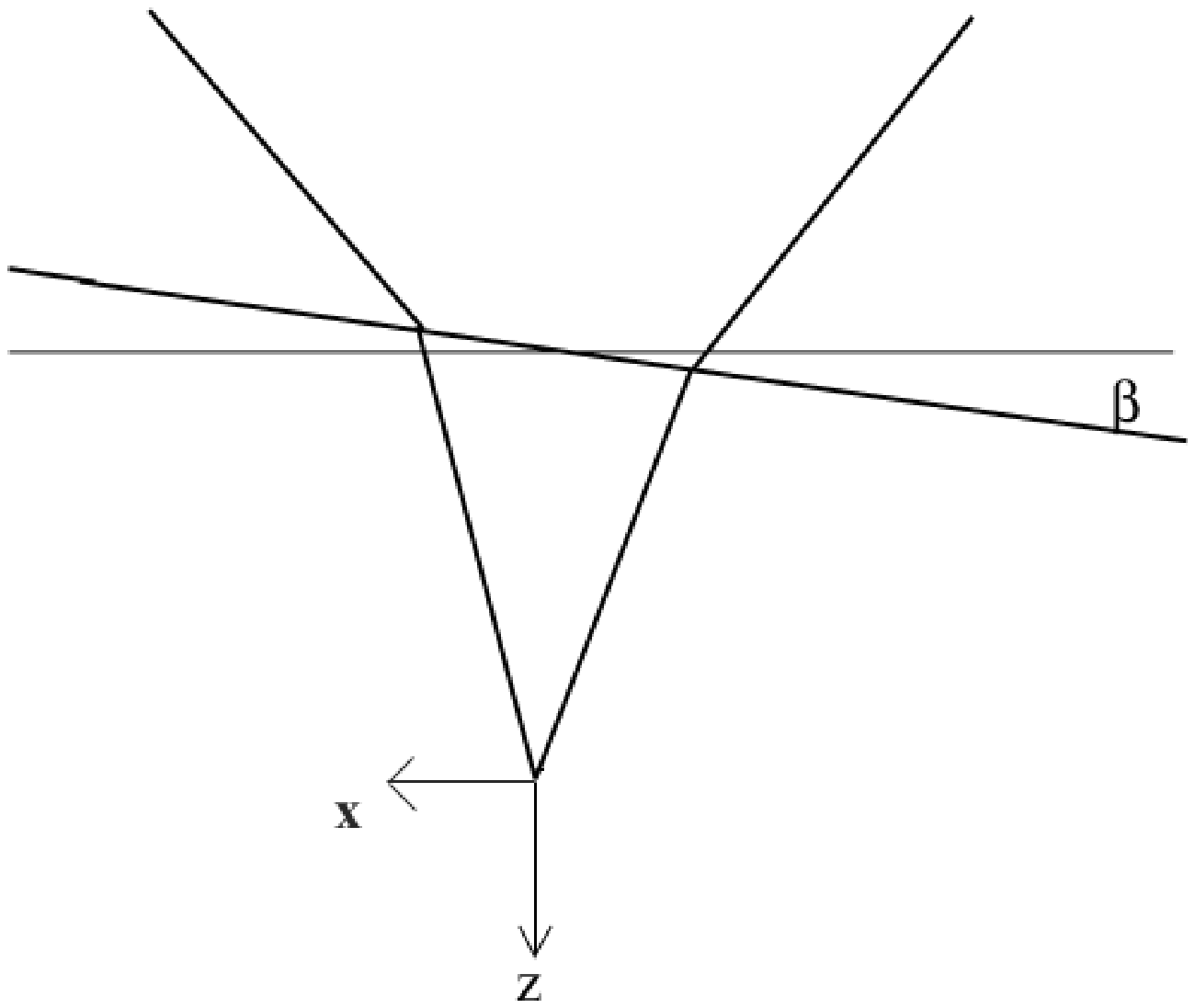}
\caption{\label{f1}  Simplified schematical drawing of the focusing geometry.}
\vspace{2cm}
\end{figure}

\end{document}